# Charge migration in organic materials: Can propagating charges affect the key physical quantities controlling their motion?

Dr. Caroline Gollub,[1,2] Dr. Stanislav Avdoshenko,[1] Dr. Rafael Gutierrez,[1] Prof. Dr. Yuri Berlin,[3] and Prof. Dr. Gianaurelio Cuniberti[1,4,5]

[1] Institute for Materials Science and Max Bergmann Center of Biomaterials, Dresden University of Technology, 01062 Dresden, Germany
[2] Max Planck Institute for the Physics of Complex Systems, 01187 Dresden, Germany
[3] Department of Chemistry, Northwestern University, Evanston, Illinois 60208-3113, USA
[4] National Center for Nanomaterials Technology, Division of IT Convergence Engineering Pohang, POSTECH, 790-784, Republic of Korea
[5] corresponding author's e-mail: *projects@nano.tu-dresden.de*

Abstract
Charge migration is a ubiquitous phenomenon with profound implications throughout many areas of chemistry, physics, biology and materials science. The long-term vision of designing functional materials with tailored molecular scale properties has triggered an increasing quest to identify prototypical systems where truly molecular conduction pathways play a fundamental role. Such pathways can be formed due to the molecular organization of various organic materials and are widely used to discuss electronic properties at the nanometer scale. Here, we present a computational methodology to study charge propagation in organic molecular stacks at nano and sub-nanoscales and exploit this methodology to demonstrate that moving charge carriers strongly affect the values of the physical quantities controlling their motion. The approach is also expected to find broad application in the field of charge migration in soft matter systems.

## 1. Introduction

With the emergence of molecular scale electronics, novel fascinating perspectives have opened for using molecules as functional building blocks of electric circuits and devices [1-4] as well as for the possibility to apply methodologies known from the study of controlled quantum dynamics [5] to tune molecular-scale device properties. This is expected to have also a strong impact on organic electronics including such areas as solar energy conversion, photonics, and sensor development [6]. For all these areas, the understanding of charge migration at nanometer length scales is crucial.

Charge motion is usually characterized by the drift mobility, which is considered as a quantity of prime importance for several applications since it determines e.g. the efficiency of charge separation in photovoltaic cells based on organic dyes, the switching speed of organic field effect transistors, and the intensity of light in light emitting diodes [7]. In addition, some basic biochemical processes in living matter are associated with charge migration at scales of several hundred angstroms. In this context, a physical picture based on electron or hole propagation is actively discussed as well [8].

It is widely accepted that charge transport processes at the molecular scale can cover a broad spectrum of possible regimes ranging from coherent propagation up to fully incoherent hopping. Contrary to most crystalline solids, the efficiency of charge migration at nano and sub-nanometer scales is strongly affected by the system's structural fluctuations as has been established, *e.g* for polymers [9], bio-molecular assemblies [10], and stacks of organic molecules [9]. According to theoretical results verified by computer simulations [11], the effect mentioned above arises if structural fluctuations cover time scales comparable with the characteristic times of charge propagation. As a result, the atomic dynamics cannot be treated as a small perturbation of an otherwise static system, but has to be included in the theoretical description of charge transport beyond standard perturbative approaches.

Different theoretical methods are meanwhile available for modeling transport properties of organic systems. For instance, classical Marcus theory [12] has been applied to compute the charge carrier mobility in various organic materials [9, 13, 14], where charge transport mainly proceeds via the mechanism of incoherent hopping. On the other hand, in the case of coherent charge transport the Landauer approach [15] combined with non-equilibrium Green function techniques is widely used to calculate electrical properties of molecular wires attached to electrodes [16-18]. The coherence breakdown resulting from the coupling of the molecular electronic subsystem to environmental degrees of freedom can be taken into account either in a phenomenological way [19] or explicitly by including the interaction with vibrational modes or with a dissipative environment [20-23]. Both cases can then be treated either with Green functions or with reduced density matrix approaches. Alternatively one can directly solve the time-dependent Schrödinger equation with appropriate boundary conditions in order to obtain the transport characteristics [24-29].

Based on these theoretical approaches, it becomes possible to identify the main factors controlling the rate and the efficiency of charge transport. Within a coarse-grained picture of the electronic structure (which will be the framework of the current study), two physical quantities naturally emerge, which provide a link between the electronic structure of molecular materials and the rate of inter-molecular charge transfer. These quantities are the electronic couplings (expressed e.g. in terms of transfer integrals), which mediate the interaction between building units making up the molecular material, and the molecular orbital (MO) energies, which characterize the electronic structure of the individual units. In what follows the two quantities defined above will be called for the sake of simplicity charge transport (CT) parameters. We note that the specific definition of the previously mentioned physical quantities will obviously depend on the degree of coarse-graining of the electronic structure.

In recent years much effort has been made to calculate both parameters for various systems at different levels of theory [30-32]. To our knowledge, most of the calculations performed so far (an exception being e.g. Ref. [27]) rely on the implicit assumption that the presence of moving charge carriers does not strongly modify the values of electronic coupling and molecular orbital energies; however, this assumption seems to break down in many situations. For example, one can expect that in the case of hole transport the coupling between two neutral organic molecules will differ from the coupling between cationic and neutral molecular species.

In the present study we present a computational methodology that is free from adjustable parameters and allows us to include in a consistent way the influence of a propagating charge onto the CT parameters of the system through which it is moving. Using this approach, we demonstrate that the explicit inclusion of a moving charge can have a dramatic effect on the values of these parameters and hence on the qualitative behavior of the transport characteristics of the system. It should be recognized that the situation considered here is distinct from changes of molecular orbital energies accompanied, for instance, by the formation of a polaronic structure [33]. In the latter case, changes in the values of the MO energies result from the polarization of the environment or from the self-trapping of a positive charge by a distortion of the molecular stack. By contrast, in the present work we address the question of whether changes in the values of molecular orbital energies and electronic couplings can occur exclusively due to the motion of charge carriers rather than caused by the interaction with the environment. In the next section we present in some detail the computational methodology developed in this work, whereas an exemplary application of the methodology proposed to the description of charge migration along a coronene molecular stack is discussed in Sec. **III**.

## 2. Computational Methodology

A key element for taking into account the interplay of conformational dynamics and charge carrier motion is the use of fractional occupation numbers [34] together with the inclusion of the electrostatic potential of net charges. The main advantage of using fractional charges is the possibility to describe partial delocalization of a moving charge over several molecular sites and thus, to go beyond a purely hopping-like propagation where the charge is fully localized at each site. Due to the strong electrostatic effects that an excess charge produces on its local environment –leading to strong shifts of the molecular orbital energies, see the ongoing discussion- the inclusion of a single electronic level per site can not catch the full complexity of the charge migration process; rather, a multi-level description will be required at this point. Before going into technical details, we will describe the general structure of our methodology, which for the sake of brevity we will call quantum dynamical charge propagation (QDCP) scheme. The approach is schematically illustrated in Figure 1.

An input geometry of the target system as obtained from a molecular dynamics (MD) trajectory (after an equilibration run) is taken as an initial structural guess. Transfer integrals and molecular orbital energies are obtained from *ab initio* electronic structure calculations. This step is called parameterization (PAR) for the sake of simplicity. The results obtained at the PAR step are exploited in building a coarse-grained Hamiltonian matrix of the system at a given time. The quantum dynamical (QD) computation of the hole density matrix determines then the new charge distribution (CD) to be included in the subsequent MD step. The resulting CD is used to mimic the electrostatic field experienced by the system. This point is essential since, as we will see in the following, charge transfer parameters do sensitively depend upon the current CD. Once a propagation cycle is completed (MD→PAR→QD→CD→MD), the Hamiltonian matrix elements (the CT parameters) and the molecular geometries are updated.

Although our approach can be exploited for molecular systems with arbitrary spatial arrangements, the details of the method will be explained for clarity using a linear molecular array as an example, specifically a stack of coronene molecules as displayed in Fig. 2. The electronic structure of each molecular building block of the stack is treated within the so-called fragment orbital (FO) approach using fractional charges as implemented e.g. in the ADF code [41] and used by other authors as well [24]. The fractional charge for a specific fragment is given by the integral of the fragment electronic density. At a given simulation time step, the influence of fractional charges of fragments which are neighbors of a given fragment $m$ (the index $m$ running over the fragments in the stack) are taken into account as classical point charges to compute the CT parameters (the PAR step in the loop of Figure 1), thus acting as an electrostatic field affecting fragment $m$. Then, the net (or grid) charge $Q_m$ (which is the fractional charge associated with the fragment) is assigned to fragment $m$ (Fig. 2). The charge distribution obtained this way affects the diagonal part of the electronic density matrix for fragment $m$, which is given by $\rho_m = \sum_j n_j |\Psi_j^m|^2$, $n_j$ being the occupation number of orbital $j$. The off-diagonal Hamiltonian matrix elements (the electronic couplings) for every pair of neighboring molecules $m$ and $n$ are also evaluated using the FO approach. For this case, we assign the net charges $Q_{mn}=Q_m+Q_n$ (see Fig. 2) to the corresponding fragment pair $m,n$. Similar to the case of the individual fragments, the environment, i.e. the surrounding molecular fragments, is mimicked by electrostatic potentials originating from point charges located at the centers of the molecules surrounding a given fragment.

It should be emphasized that the notations fractional charge and net charge only highlight the two different ways charges are taken into account in our computational approach: On the PAR step we have fractional charges determined *ab initio* on each fragment while on the MD step the term net charges is used to denote classical point charges placed at the center of each fragment.

Within the PAR step of the closed-loop propagation scheme a coarse-grained electronic Hamiltonian is set up as:

$$H = \sum_{m=1}^{M} \sum_{i=0}^{L} \left\{ \varepsilon_i^m(t)(a_i^m)^+ a_i^m - \sum_{n=m+1}^{M} \sum_{j=i+1}^{L} T_{ij}^{mn}(a_i^m)^+ a_j^n \right\} \qquad (1)$$

where $\varepsilon_i^m(t) = \langle \Phi_i^m(t) | H | \Phi_i^m(t) \rangle$ and $T_{ij}^{mn}(t) = \langle \Phi_i^m(t) | H | \Phi_j^n(t) \rangle$. The wave functions $\Phi_j^n(t)$ are MOs of the individual fragments at the simulation time $t$, and the operators $a_i^{m\dagger}$ and $a_i^m$ create and annihilate an electronic excitation on fragment $m$ and on molecular orbital $i$, respectively. The sum over the orbital indices $i$ and $j$ runs only over occupied states. The total wave function can then be expressed as a linear combination of the $l$-th occupied molecular orbital at each fragment $m$:

$$\Psi(t) = \sum_{m=1}^{M} \sum_{l=0}^{L} c_l^m(t) c_l^m \Phi_l^m. \qquad (2)$$

The orbital index $l=0$ refers to the HOMO orbital. The lower orbitals with $l \geq 1$ were denoted by HOMO-$l$. In the special case of coronene, seven levels have been taken into account per site.

The choice of a multilevel approach to the problem of charge carrier dynamics is motivated by two main facts: Firstly, at finite temperatures the values of HOMO-HOMO-$l$ electronic coupling elements of neighboring molecules are close to the value of HOMO-HOMO coupling. Secondly, charges involved in the transport modify the orbital energies, thus affecting the effective coupling between MOs at neighboring fragments. Consequently, the manifold of HOMO-$l$ levels ($l \geq 1$) may also get involved in the transport depending on the obtained CT parameters. In all calculations, we assume that the fractional hole population undergoes a fast transition with a characteristic rate $\gamma_{intra}$ from HOMO-$l$ to the HOMO level. Further, an additional molecular unit is added at the edge of the stack, which irreversibly traps a charge carrier (similarly, for a higher dimensional arrangement a set of traps should be included in the simulation). The main function of this trapping site is to avoid unphysical multiple reflections over the simulation time which result from the short length of the studied molecular stacks.

The hole time evolution was computed according to the dissipative Liouville-von Neumann equation [35] for the temporal evolution of the quantum system's density matrix $\rho(t)$:

$$\dot{\rho}(t) = L\rho(t) = -i[H, \rho(t)] + L_D(\rho(t)). \tag{3}$$

The intra-fragment transition processes (characterized by $\gamma_{intra}$) as well as the trapping effects at the terminal site (characterized by a parameter $\gamma_{trap}$) can be included in the dissipative kernel $L_D$ written in the Lindblad form [36]

$$L_D(\rho(t)) = \sum_{i=0} C_i \rho(t) C_i^\dagger - \frac{1}{2}[C_i^\dagger C_i, \rho(t)]_+,$$

where $C_i = \gamma |a\rangle\langle b|$, $C_i^\dagger = \gamma |b\rangle\langle a|$ are the Lindblad operators corresponding to raising and lowering operators of the $i$-th two-level system, and $\gamma$ represents either $\gamma_{intra}$ or $\gamma_{trap}$.

The formal solution of Eq. 3 for the initial density matrix $\rho(t')$ and step $\Delta t = t'' - t'$ is given by

$$\rho(t'') = e^{L(\Delta t)} \rho(t').$$

For the numerical implementation, the Faber polynomial method [35] was applied to approximate the exponential of the matrix $L$. The real-time propagation was performed for a time interval of 1 fs with the time step $\Delta t = 0.1$ fs. A linear interpolation was used to obtain the Hamiltonian matrix from the parameterization part for every 0.1 fs.

Finally, the QDCP cycle closes with an MD step following the QD propagation step of the charge carrier wave function. After the time propagation of the density matrix the new geometry of the molecular step is evaluated based on the previous atomic geometry (in the first step based on the geometry taken from the equilibration) and its corresponding velocities using MD simulations. The actual charge distribution obtained in the previous step from the QD calculations is approximated by Gaussian blur distributions of point charges located at the center of the fragments (e.g. a coronene molecule). The MD propagation time is also equal to 1.0 fs, as in case of the QD calculations, but in this case the integration step is taken to be 0.5 fs. The resulting geometry obtained within this MD step is then used in the next cycle of the QDCP calculation together with the new charge distribution from the QD result to calculate new CT parameters and continue the closed-loop propagation scheme.

## 3. Results and Discussion

**3.1 Electronic structure and dynamics in coronene molecular stacks** The potential of the *ab-initio* QDCP propagation scheme described in Section II is demonstrated by applying this scheme to the calculation of the mobility of a positive charge carrier (hole) along a π-π stack of coronene molecules analogous to those studied experimentally by Feng *et al.* [37]. Note that our approach is not limited to the case of hole migration, but can equally be applicable to the transport of electrons. Using the approach discussed above, we show that the mobility of holes strongly depends on the temperature in accordance with the experimental results [37]. To advance one of our main results, we reveal a quasi ballistic behavior of charge carriers at short times as evidenced by a steep linear increase of their diffusion coefficient $D(t)$ with time in the fs range. This behavior, however, transforms very rapidly into a diffusive regime of hole motion that manifests itself as a quasi-saturation of $D(t)$ as a function of time. The latter regime dominates the transport characteristics on longer (about a few 100 fs) time scales.

Initially, an equilibration of the ideal stack of coronene molecules with an inter-molecular distance of 3.6 Å [38] was performed with the SCC-DFTB [39] methodology as implemented in the DFTB+ package. This method, being based on a density-functional parameterization of a tight-binding Hamiltonian is computationally very efficient and it has been used in the past years to address a very broad variety of issues [40]. An integration step of 0.5 fs was used during the equilibration and environmental effects were introduced by coupling the system to the Andersen thermostat [41]. After the equilibration run, a geometry was taken as an initial guess for the QDCP charge propagation. In the following step, the parameterization of the molecular orbital energies and nearest-neighbor transfer integrals, see Eq. 1, at the *ab initio* level of theory was performed with the ADF package [42] in order to obtain the required quantities for solving the coarse-grained charge carrier dynamics. The ADF package is chosen for the electronic structure calculations as it provides the possibility to take fractional charges into account. In every QDCP cycle, the MO energies of each fragment of the system are determined using density functional theory with semi-local GGA type PBE exchange correlation functional [43] and double zeta polarized basis set. The fractional charges (for $q_m$>0.01e) of neighboring molecules were taken into account as point charges and the corresponding net charge $Q_m$ is assigned to the fragment (see also Fig. 2 for reference). The characteristic transition rate $\gamma_{intra}$ from HOMO-$l$ (with $l$=6) to the HOMO levels was taken to be 0.1 fs$^{-1}$.

To calculate the mobility of a hole in the system under investigation, we approximate the molecular stack of coronene molecules by a one dimensional chain, in which each molecule can be considered as an individual fragment (see Sec. **II**). The successful application of our approach to model systems is based on a successive calculation of the time evolution of the charge carrier wave function in the case where nuclear and electronic degrees of freedom are coupled. Currently, full *ab initio* Ehrenfest dynamics [44] in the basis of atomic orbitals at the attosecond scale is considered as a powerful theoretical tool for the study of the electronic density response of single molecules to external time-dependent potentials. In the theoretical analysis of transport/transfer problems the evolution of the charge carrier density can be treated as the external time-dependent potential in the space of the fragments MOs. However, in organic materials a weak MO coupling leads to longer characteristic times of transitions between sites which typically occur on the time scale of tens of femtoseconds. Therefore this allows for an increased propagation step. The duration of each propagation cycle was taken to be 1 fs to cover the dynamical fluctuations of the MO energies and transfer integrals. In the present QDCP scheme the electronic structure was reduced to the active electronic space (for holes, the highest occupied MOs) involved in the transport and represented, for the quantum dynamical propagation, in a coarse-grained MO basis. In order to describe the charge distribution in the considered time step, corresponding fractional net charges $Q_m$ of a given fragment $m$ and point charges $q_m$ mimicking the effective field of neighbors were included in the calculations of charge transport parameters (Fig. 2). The energy profile obtained for the hole transport is also shown in the inset of Fig. 2.

From the QD propagation a new charge carrier distribution was obtained, which was included in the subsequent MD calculation as molecular mechanics (MM) potentials of a QM/MM calculation, performed with the DFTB code [39]. As a consequence, the structural dynamics will be affected by the updated charge redistribution from the quantum dynamical calculation. After a full cycle of the QDCP propagation a new charge distribution and a new geometry were obtained for the next step. In the chosen model system, the coronene stack involved ten molecules, which proved to be sufficient to model charge transport in the studied system. This is basically related to the fact that the charge wave function was extended at each simulation time step over no more than three sites, so that the system length (thus about 3 times larger than the width of the wave function) was enough to get reliable results.

As follows from our computations, after a time period of $t=150$ fs the charge is spread mainly over the first three fragments of the coronene stack. This leads to a decrease of the HOMO energies since the fragment populations become lower. According to the data shown on the inset of Fig. 2, the energy difference between fully and partially occupied orbitals (sites 1 and 10) is on the order of 1 eV. This is much larger than the thermal fluctuations of the molecular orbital energies (~0.1 eV). The behavior of the MOs displayed in the inset of Fig. 2 was then mimic within the MD step of the loop of Figure 1 by applying a classical electrostatic potential, see Fig. 3a.

Using these results, we also obtain the time behavior of the total difference in electron density for the neutral and charged systems as shown in Fig. 3c, where a snapshot of the hole behavior at $t=150$ fs is shown. According to the calculated electron distribution, HOMO and LUMO states of the complete system localize in space as shown in Fig. 3b.

The QDCP propagation scheme was applied to the coronene stack coupling to a thermostat with a certain temperature $T$. The data deduced from the QDCP calculations at different temperatures in the range from 300 K up to 500 K in steps of 50 K are shown in Fig. 4. When comparing the two extreme cases of low (300 K) and high (500 K) temperatures, we see that the wave packet can propagate coherently across the stack, while it splits up at higher temperatures and this is accompanied by a rapid broadening of the charge distribution. Hence, at room temperature it turns out that the dynamical structural disorder cannot overcome the collective stabilization of the wave packet resulting from the interaction with nearest-neighboring and next-to-nearest-neighboring sites. MD simulations provide additional evidence in favor of the validity of this conclusion. As the temperature increases (up to 400 K) the molecular orbital energies and transfer integrals are increasingly affected by the dynamical structural disorder in the stack; this in turn leads to a broadening of the wave packet and to a destruction of the collective stabilization. Finally, at high temperatures (500 K) the strong dynamical disorder results in a splitting of the wave packet and partial localization.

**3.2 Diffusion coefficient and charge carrier mobility**
The QDCP propagation scheme also enables us to obtain observables that characterize the charge carrier transport. One of these observables is the diffusion coefficient of holes, *D*, plotted in Fig. 5 (black solid line) at *T*=300 K as a function of time. According to the data presented in this figure the time behavior of the diffusion coefficient differs in different time intervals. For the first few femtoseconds, the large diffusion coefficient reflects the initial delocalization of the charge carrier due to a quasi-ballistic motion as follows from the initial steep linear increase of the diffusion coefficient with time. This is related to the localization of the hole on a single site at *t*=0. Later, however, the motion of the charge carrier becomes diffusive. As a result the *D* values drop down and do not change in the range from 150 fs to 625 fs (for the diffusive-like regime, see shaded region in Fig. 5). The long-time tail of the time-dependence found for *D*(*t*) can be associated with the irreversible localization of the wave packet on the last site serving as a trap of the moving charge. It is remarkable that in the limit where a charge does not affect the values of MO energies and transfer integrals, the diffusion coefficient exhibits a completely different temporal behavior. This is evident from the comparison of solid and dashed lines in Fig. 5, the latter obtained *without* the influence of charge carriers on the parameters controlling their motion. In the case where the impact of moving carriers on the values of the CT parameters is neglected -as it happens in most of the approaches dealing with charge transport- the diffusive regime does not set in within the time scale studied (dashed line in Fig. 5); instead, only a ballistic behavior with *D*(*t*)~*t* is obtained. As a result, we expect a strong overestimation of the charge carrier mobility if the influence of the propagating charge onto the CT parameters is not explicitly taken into account. Based on these findings we conclude that a moving charge can affect its own motion in a self-consistent way. We emphasize at this point the main difference between the approach presented in this study and other standard techniques which solve a time-dependent problem (either using the Schrödinger equation or the density matrix formalism) to study charge migration [24-26,28-30]: In the latter methodologies no influence of the charge carrier on the CT parameters is considered. As a result, the Hamiltonian matrix used as input for the time propagation does not need to be updated at each time step to include the changes in the electronic structure induced by the moving charge. In other words, in standard approaches we have an input geometry, a parameterization step (coarse-graining) and then the computation of the charge density time evolution from where the required transport observables can be obtained. In more complex situations where structural fluctuations play a dominant role, an additional MD simulation is performed and the previously mentioned steps are carried out at each snapshot along the MD trajectory. In either case, the time propagation and the computation of the CT parameters can be decoupled from each other. In contrast, within the current approach, molecular dynamics, parameterization, and time propagation steps are coupled and build a single loop. Hence, our methodology reduces to the standard approach as the excess charge becomes equal zero at the parameterization step.

Using the computed diffusion coefficients $D(t)$ in the range from 70 fs to 300 fs together with the Einstein relation $\mu(t,T)= D(t)/k_B T$, we can estimate the charge carrier mobility $\mu$ for different temperatures. The results are shown in Fig. 6. As follows from the upper panel of this figure, the hole mobility increases as the temperature rises up to 400 K in accordance with experiments in which the temperature dependence of the mobility was measured for similar systems [36].

This behavior can be understood since for higher temperatures the thermal fluctuations allow for a more effective coupling along the stack, thus leading to larger $\mu$ values. Moreover, the QDCP calculations predict that the dependence $\mu$ vs. $T$ has a maximum between 400 K and 450 K. For $T>450$ K, however, the influence of thermal fluctuations becomes detrimental for charge transport: a thermally induced increase of the inter-molecular distances in the coronene stack leads to a weakening of the π-π coupling and thus to a reduction of the mobility. The latter effect is consistent with the experimentally observed sublimation region [45]. Although it is difficult to clearly distinguish the crossover between different transport regimes, ballistic and diffusive-like mechanisms can be described by the QDCP propagation scheme (see Fig. 6, lower panel). In particular, at 450 K a transition from the initial ballistic regime to diffusive-like transport regime occurs after 70 fs, whereas at e.g. 350 K the crossover takes place already after 25 fs. We emphasize that our approach based on the QDCP propagation methodology can equally be applied to other π-π stacked systems with transport properties which have been studied experimentally (see e.g. Refs. 46 and 47).

## 4. Conclusions

In summary, we have developed a computational methodology that allows us to quantify charge transport in molecular materials via a quantum dynamical propagation of the carrier wave function. The proposed methodology takes into account the influence of the charge on the physical quantities controlling its propagation. Our approach provides a common basis for studying charge transport on different length and time scales as well as in different systems including single molecules, biomolecular assemblies and stacks of organic molecules. The essential feature of our method is the influence of the moving charge on the electronic coupling matrix elements and molecular orbital energies which are considered to be the main physical quantities governing charge motion. Due to this influence the moving charge can affect its own motion in a self-consistent way. Using the QDCP computational, the experimentally measured hole mobility on a coronene stack as well as its temperature dependence were reproduced. This enables us to expect that the proposed methodology will have a strong impact in understanding the mechanisms controlling charge migration at the nanoscale and that it will help in the rational design of molecular organic materials with optimized charge transport properties.

**Acknowledgment**

The authors thank Denis Andrienko for fruitful discussions. This work was partially funded by the Max Planck Institute for the Physics of Complex Systems, the Erasmus Mundus Program External Co-operation (EM ECW-L04 TUD 08-11), the Erasmus Mundus Program in Nanoscience and Nanotechnology, the Cluster of Excellence ECEMP, ``European Centre for Emerging Materials and Processes'' within the excellence initiative of the Free State of Saxony, and the South Korea Ministry of Education, Science, and Technology Program ``World Class University'' under Contract No. R31-2008-000-10100-0. Computational resources were provided by the Center for Information Services and High Performance Computing (ZIH) of Dresden University of Technology.

**Figure captions**

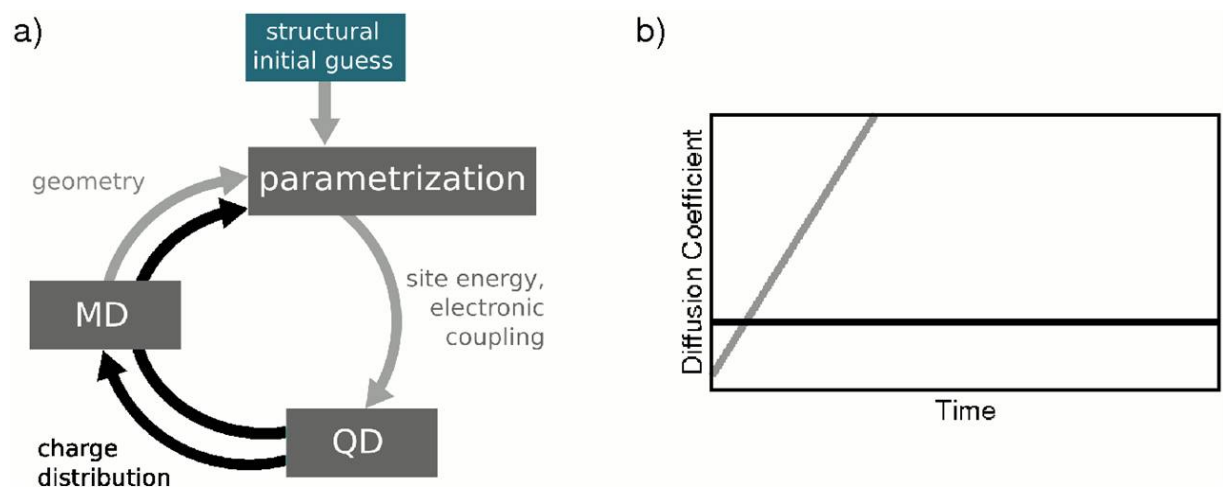

**Figure 1:** The QDCP methodology**.** *(a) The motion of a charge through a material is monitored in time and space by exploiting successively ab initio parameterization tools, quantum dynamics and molecular dynamics in form of a closed loop propagation. As an essential new feature, the effect of the moving charges is consistently taken into account in the quantum mechanical evaluation of the electronic transfer integrals and molecular orbital energies, as well as in the molecular dynamics trajectory (black arrows). If the influence of the propagating charges onto the material CT parameters is included (black line in panel b), D(t) becomes nearly constant after very short transient times, thus indicating diffusive charge motion. In contrast, if the influence of the propagating charge onto the electronic structure of the underlying system is not taken into account (gray line in panel b), the charge motion remains ballistic (linear time-dependence of D(t)) over a broad time window. As a consequence, the diffusion coefficient and hence the mobility of the charge carriers can be strongly overestimated.*

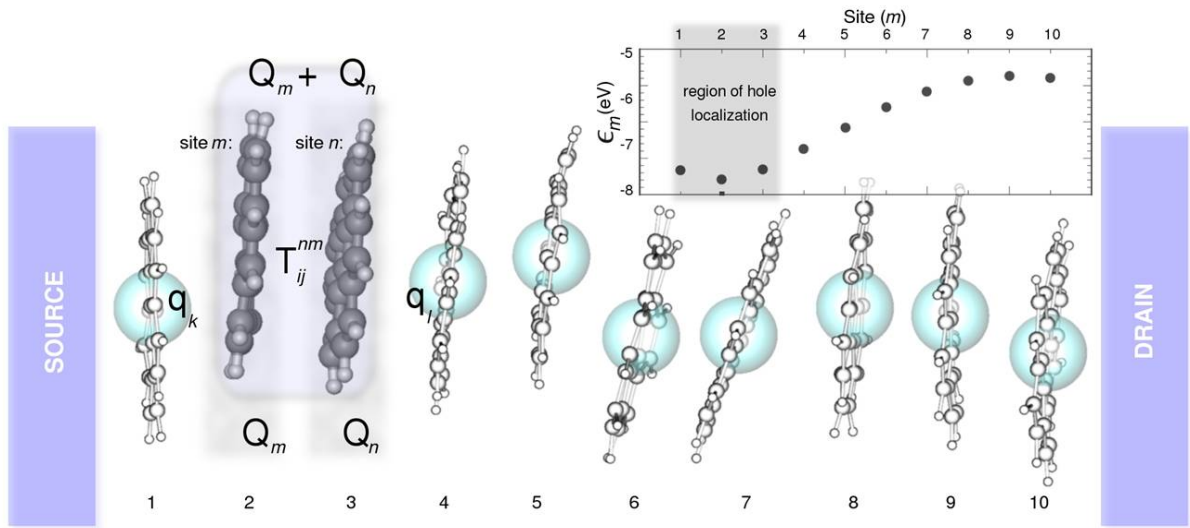

**Figure 2:** System setup. *To benchmark the QDCP method a one-dimensional model system consisting of ten coronene molecules has been chosen. Source and drain reflect the boundary conditions applied in the QDCP propagation. Point charges $q_m$ and net charges $Q_m$ were included in the method to account adequately for the influence of the moving charge carriers on the values of charge transport parameters of the molecular material. As a result, the charge transfer properties - molecular orbital energies ($\varepsilon_i^m$) and electronic couplings between the orbitals of neighboring molecules ($T_{ij}^{mn}$) - evaluated in the presence of the charge carriers strongly differ from corresponding neutral calculations. This is illustrated in the inset which displays the orbital energy profile for hole motion at a certain time point (t=150 fs). The excess charge is mainly spread over the first three molecules (gray region); however, it also influences the molecular orbital energies of the neighbors up to site 7.*

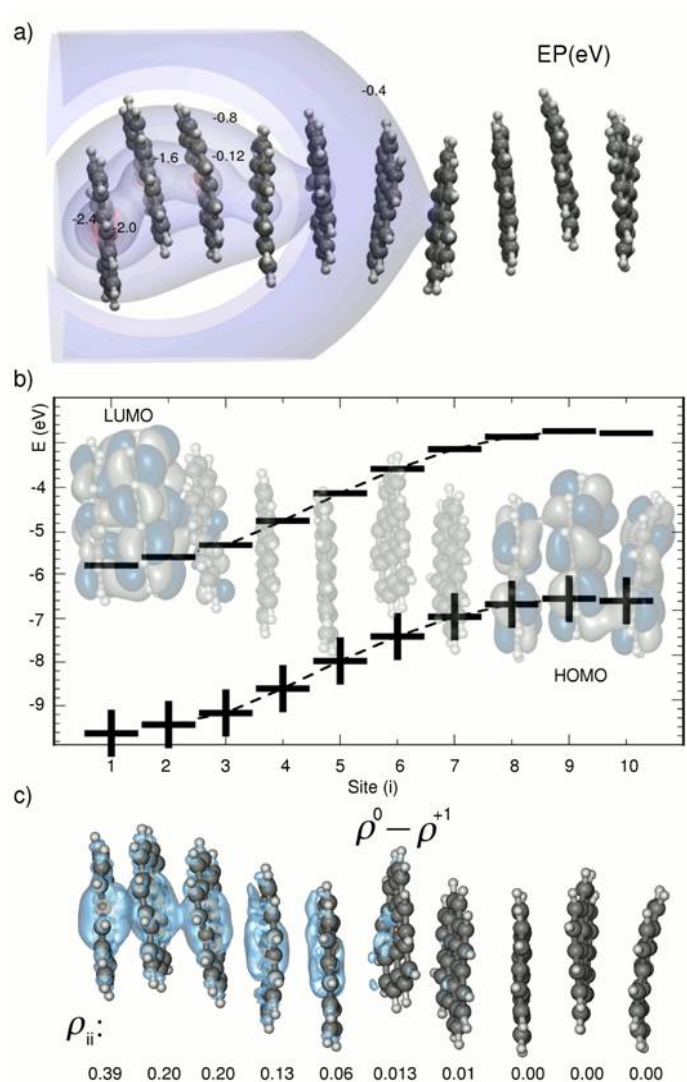

**Figure 3:** Charge migration in the stack of organic molecules. *a) The electrostatic potential (EP) determined for an equivalent charge distribution as in the inset of Fig. 2 at the time step t=150 fs clearly indicates the decisive role of the charge carriers explicitly included in the calculations. As a consequence of the EP and the energetic profile of the molecular orbital energies adapting to the charge distribution, the HOMO and LUMO molecular orbitals (obtained from QM/MM calculations) of the stack localize as shown in panel b). The QDCP calculations enable one to follow the real-space distribution of the density of the charge carriers. In panel c) the hole density distribution is shown for the specified populations per site (bottom line). It is expressed as the difference in electron densities for the neutral $\rho^0$ and charged $\rho^{+1}$ system.*

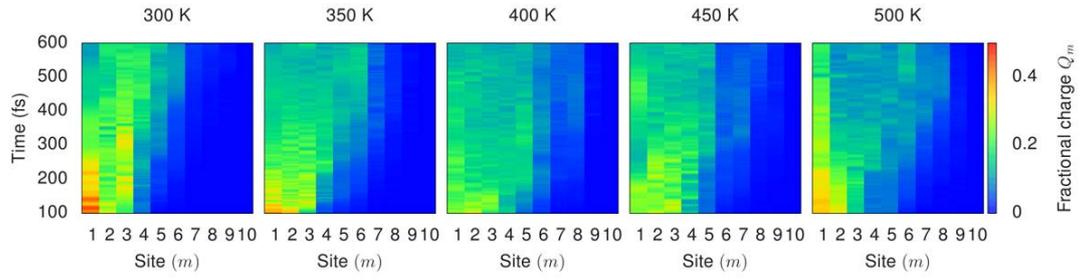

**Figure 4:** QDCP charge dynamics. *The quantum time evolution of the charge carriers under the influence of the atomic dynamics is visualized as a function of time and of the populated molecular sites. The QDCP charge propagation was performed for different temperatures from 300 K to 500 K in steps of 50 K. Note the rapid broadening of the charge distribution along the stack when the temperature increases.*

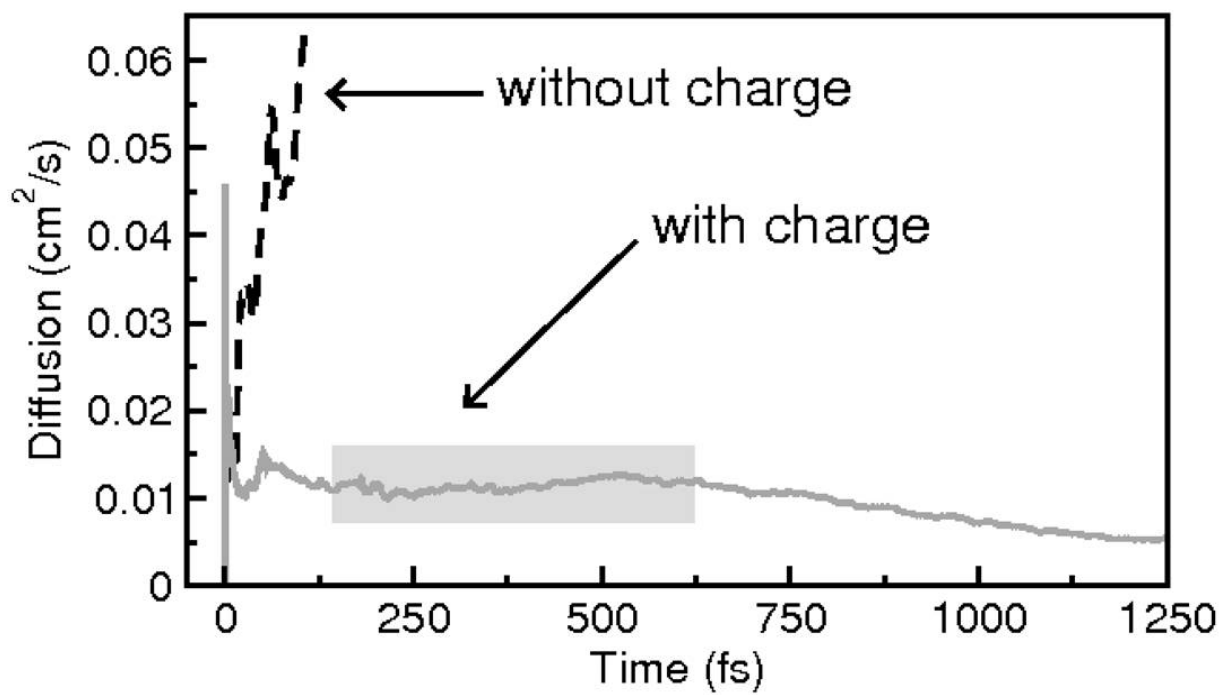

**Figure 5:** Time-dependent diffusion coefficient calculated with the QDCP method. *The migration of a hole through a stack of coronene molecules modeled with the QDCP method can be quantified in terms of the diffusion coefficient. A corresponding result for 300 K is shown by the gray solid line. After an initial increase, a time window (shaded box, from 170 fs to 600 fs) is reached where the diffusion coefficient does not vary significantly with time indicating the onset of a diffusive-like transport regime. Using the same molecular system at 300 K, but neglecting the explicit effect of the charge carriers on the values of the charge transport parameters leads to a ballistic-like behavior of the diffusion coefficient (black dashed line) and thus to a considerable overestimation of the mobility.*

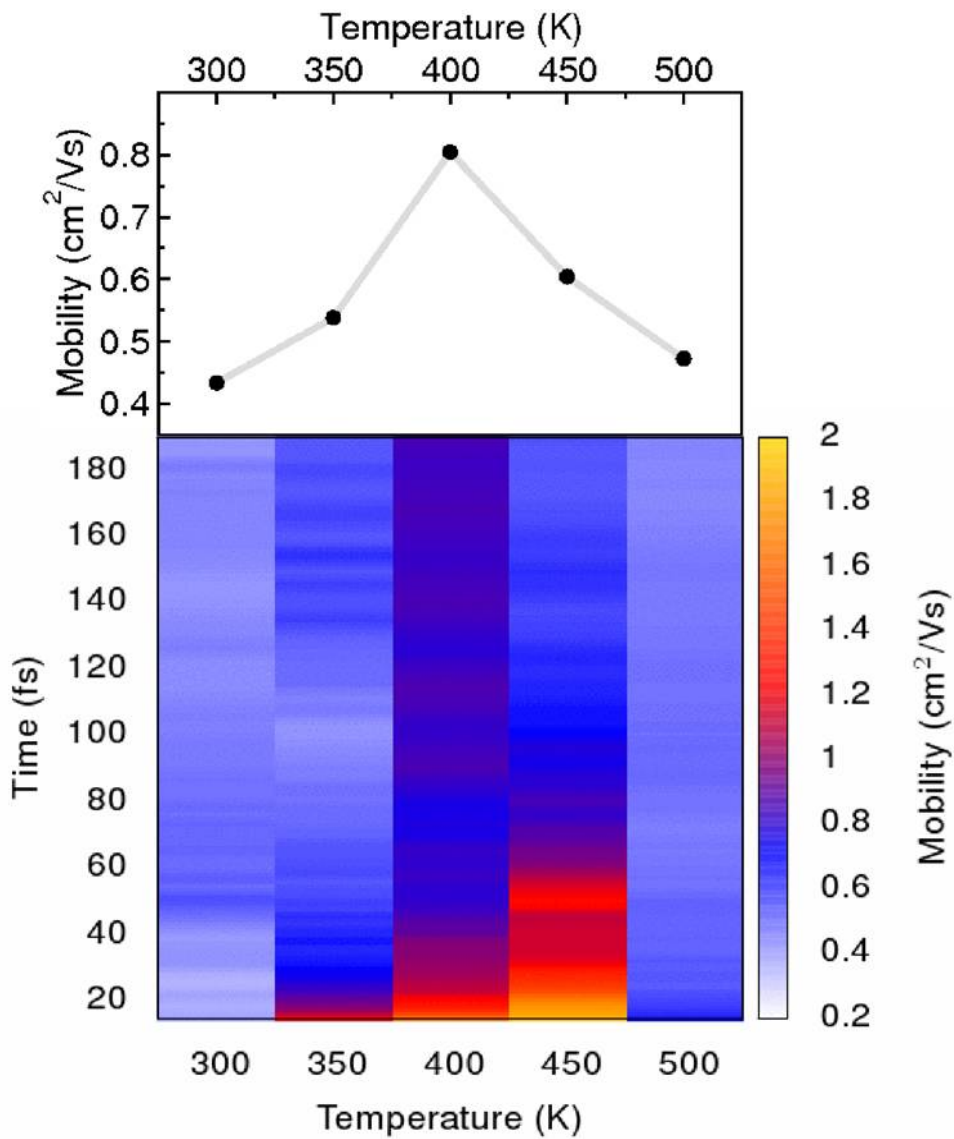

**Figure 6:** Hole mobility along the stack of coronene molecules vs. temperature and time. *From the diffusion coefficient the mobility of the charge carriers can be evaluated using the Einstein relation. In the upper panel the calculated hole mobility across the coronene stack is shown as a function of the temperature. In each case, the nearly time-independent diffusion coefficient was used, as indicated in Fig. 5 by the shaded box. In the lower panel the time-resolved hole mobility is depicted for the same temperature range starting from 20 fs after the hole was localized on the first coronene molecule. The behavior shown partly reflects the fact that two types of mechanisms interfere at these early times in the transport process. A transition from a ballistic to a diffusive regime can be found at 50 fs (70 fs) for T=400 K (T=450 K).*